\documentclass[11pt,a4paper]{article}

\usepackage{amsmath,amssymb}
\usepackage{graphicx}
\usepackage{bm}
\usepackage{cite}
\usepackage{geometry}
\usepackage{hyperref}
\usepackage{xcolor}
\usepackage{xcolor}
\colorlet{red}{black}

\newcommand{\be}{\begin{equation}}
\newcommand{\ee}{\end{equation}}
\newcommand{\bea}{\begin{eqnarray}}
\newcommand{\eea}{\end{eqnarray}}
\begin{document}

\title{Toward a Unified Axion Cosmology}

\author{
Takeshi Fukuyama\\[2mm]
{\small Research Center for Nuclear Physics (RCNP)}\\
{\small Osaka University, Ibaraki, Osaka 567--0047, Japan}
}
\date{}

\maketitle

\begin{abstract}
We investigate a unified axion framework connecting grand unification,
string-inspired axion-like particles, and cosmology.  The QCD axion is
identified with the relative phase along the $D$-flat direction of the
minimal supersymmetric $SO(10)$ grand unified theory, with a
Peccei--Quinn scale of order $10^{12}$ GeV, while heterotic string
compactification provides a broad spectrum of axion-like particles
whose masses are exponentially controlled by nonperturbative instanton
actions.  \textcolor{red}{
Within this multi-axion structure, cosmological requirements select
two qualitatively different directions in axion field space.
A light massive eigenmode characterized by a scale of order
$10^{-22}\,{\rm eV}$ is assumed to constitute the dominant
dark-matter component and can form a Bose--Einstein condensate whose
attractive self-interaction leads to nonlinear collapse, providing a
possible mechanism for the early formation of massive seeds of
high-redshift supermassive black holes.
A second, aligned collective direction has a characteristic
dynamical scale of order $10^{-28}\,{\rm eV}$ and becomes relevant
near the matter--radiation transition, where it can transiently
increase the pre-recombination expansion rate and thereby reduce the
sound horizon.
} An order-of-magnitude estimate shows that addressing the Hubble tension
requires a sound-horizon-weighted axion fraction of order $0.1$,
corresponding, for a large initial displacement, to an effective decay
constant of order $10^{18}$ GeV, substantially larger than the
microscopic heterotic scale of order $10^{16}$ GeV.  We show that a
multi-axion--multi-instanton system can in principle provide
\textcolor{red}{an aligned collective direction with such an enhanced
effective field range, together with a multi-harmonic potential
allowing rapid post-critical dilution.}
Thus cosmological requirements do not merely select
\textcolor{red}{characteristic ultralight scales and cosmologically
relevant directions from the string axiverse, but also impose
nontrivial constraints on the alignment and harmonic structure of
the underlying multi-axion sector.}
\end{abstract}

\section{Introduction}

The Peccei--Quinn (PQ) mechanism provides a compelling
solution to the strong CP problem in quantum
chromodynamics (QCD) \cite{PQ,Weinberg,Wilczek}.
Its spontaneous symmetry breaking predicts the existence
of a pseudo Nambu--Goldstone boson, the QCD axion,
whose mass is generated by non-perturbative QCD effects.
The invisible axion models of
Kim--Shifman--Vainshtein--Zakharov (KSVZ) and
Dine--Fischler--Srednicki--Zhitnitsky (DFSZ)
provide phenomenologically viable realizations
\cite{Kim,SVZ,DFSZ,Zhitnitsky},
and the QCD axion has long been regarded as
a promising dark matter candidate.

The QCD axion, however, need not be the only axionic
degree of freedom relevant to cosmology.
Two particularly important cosmological issues
motivate us to investigate a broader axion sector.
One is the existence of supermassive black holes
(SMBHs) with masses of order $10^9 M_\odot$
already at redshifts $z\gtrsim7$,
which places severe constraints on their formation
and growth.
The other is the Hubble tension, namely the
persistent discrepancy between the Hubble constant
inferred from the cosmic microwave background
within the standard $\Lambda$CDM framework
and that measured by local distance indicators
\cite{Planck,Riess}.

From the theoretical point of view,
the QCD axion itself need not originate
from string compactification.
In our previous work \cite{FukuSO10},
we have shown that within the minimal
supersymmetric $SO(10)$ grand unified theory
the QCD axion can be identified with the
physical relative phase along the $D$-flat
direction of the $\mathbf{126}$ and
$\overline{\mathbf{126}}$ Higgs multiplets.
The corresponding PQ scale is related to
the intermediate Pati--Salam symmetry-breaking
scale, leading to $f_a\sim10^{12}$ GeV.

On the other hand, heterotic string compactification
provides a theoretically motivated framework
for a much broader axion sector.
Besides the model-independent axion,
compactification gives rise to numerous
model-dependent axions associated with
the topology of the internal manifold
\cite{Svrcek,Axiverse,Marsh}.
Their characteristic microscopic decay constants
can be of order $f_i\sim10^{16}$ GeV,
while their non-perturbative potential scales
depend exponentially on the corresponding
instanton actions.
Consequently, a broad spectrum of ultralight
axion-like particles (ALPs) can arise.

It is noteworthy that the bottom-up approach
based on supersymmetric grand unified theory
(GUT) and the top-down approach based on
heterotic string theory point toward comparable
energy scales.
While gauge coupling unification gives
$M_{\rm GUT}\simeq2.0\times10^{16}$ GeV,
the characteristic heterotic axion decay
constant is of order $10^{16}$ GeV.
Although this numerical proximity does not
by itself establish a common microscopic
origin, it motivates us to consider grand
unification and the string axiverse within
a unified high-energy framework.

Throughout this paper, we distinguish the
QCD axion responsible for the solution of
the strong CP problem from the string-inspired
ultralight ALPs relevant to cosmology.

An essential point of the present work is
that the cosmologically relevant degrees
of freedom need not coincide with individual
microscopic axions.
In a multi-axion--multi-instanton system,
different cosmological phenomena can probe
different physical directions of the same
multidimensional axion field space.

We focus on two such cosmological directions.
The first is a light massive direction
$\phi_{22}$ characterized by
\be
m_{22}=O(10^{-22})\,{\rm eV},
\ee
which is assumed to constitute the dominant
dark-matter component.
Its large occupation number and attractive
self-interaction are relevant to our
Bose--Einstein condensation (BEC) scenario
for the early formation of supermassive
black holes \cite{FMT,FukuyamaSMBH}.

The second is a collective direction
$\phi_{28}$ with a characteristic
dynamical scale
\be
m_{28}^{\rm char}=O(10^{-28})\,{\rm eV}.
\ee
Its evolution becomes important near
the matter--radiation transition,
where it can transiently increase the
pre-recombination expansion rate and
thereby reduce the sound horizon
\cite{PoulinEDE,KarwalKamionkowski}.

These two cosmological directions require
not only different characteristic energy
scales but also qualitatively different
scalar potentials.
The dark-matter direction is described
by an ordinary axion potential,
\be
V_{22}=\Lambda_{22}^4
\left(1-\cos\frac{\phi_{22}}{f_{22}}\right),
\ee
with nonvanishing quadratic curvature and
an attractive quartic self-interaction.

In contrast, the early-dark-energy component
must dilute sufficiently rapidly after
the onset of coherent oscillations.
A representative potential satisfying
this requirement is
\be
V_{28}=\Lambda_{28}^4
\left(1-\cos\frac{\phi_{28}}{f_{\rm eff}}
\right)^3.
\ee
At its minimum, the quadratic and quartic
derivatives vanish while the sixth-order
derivative remains nonzero.
Consequently, $m_{28}^{\rm char}$ should
not be interpreted as the small-oscillation
mass of an individual microscopic axion.

Furthermore, the cosmological analysis
in Sec.~3 gives the representative scales
\be
f_{22}\sim10^{16}\ {\rm GeV},
\qquad
f_{\rm eff}\sim10^{18}\ {\rm GeV},
\ee
together with the hierarchy
\be
\frac{\Lambda_{22}^4}{\Lambda_{28}^4}
\sim10^8.
\ee
Thus cosmology imposes simultaneous
requirements on the field-space directions,
the functional forms of their potentials,
and their non-perturbative energy scales.

The central question of the present paper
is whether these requirements can be
accommodated within a common heterotic
compactification.
The existence of a broad axion mass spectrum
alone is not sufficient to answer this
question, since the axion decay constants,
instanton actions, and instanton charge
vectors cannot in general be regarded
as independent parameters.

In heterotic compactification, both the
axion kinetic matrix and the worldsheet-instanton
actions are determined by the geometry
of the same internal manifold.
We examine the geometrical constraint
relating the instanton action to the
decay constant and its implications for
the two cosmological directions.
In particular, the simultaneous requirement
of a large instanton action and an enhanced
effective decay constant motivates
alignment in the multi-axion field space
\cite{KNP,ChoiKimYun,HigakiTakahashi,Ali}.

A further question concerns the origin
of the two qualitatively different
potential shapes.
The suppression of the quadratic and
quartic terms in the early-dark-energy
potential requires specific relations
among the non-perturbative contributions.

We investigate this problem within an
effective supergravity construction,
in which first-order and third-order
holomorphic zeros generate quadratic
and sixth-order scalar potentials,
respectively.
This formulation also relates the hierarchy
of the two potential scales to the
parameters of the effective superpotential
and the vacuum expectation values of
the K\"ahler moduli.

The purpose of the present work is
therefore not merely to select two
ultralight energy scales from the
string axiverse, but to translate
the cosmological requirements into
geometrical and holomorphic constraints
on the underlying axion sector.

Although the simultaneous realization
of these conditions in a specific
heterotic compactification remains
an open problem, the present analysis
identifies concrete requirements on
the internal geometry, the non-perturbative
superpotential, and moduli stabilization.
In this sense, cosmology can serve as
a selection principle for string
compactification.

The paper is organized as follows.
In Sec.~2 we discuss the QCD axion in
the minimal supersymmetric $SO(10)$ GUT
and the string-inspired ALP sector.
In Sec.~3 we investigate the cosmological
requirements associated with dark matter,
BEC-induced SMBH formation, and the
Hubble tension.
In Sec.~4 we translate these requirements
into constraints on heterotic compactification,
focusing on the geometrical relation between
instanton actions and axion decay constants,
the alignment of collective directions,
and the holomorphic structure of the
effective superpotential.
Section~5 is devoted to discussion
and conclusions.
\section{QCD axion and the string-inspired axion-like particles}

\subsection{QCD axion}

We begin with the conventional QCD axion, which was originally
introduced as the pseudo Nambu--Goldstone boson associated with the
spontaneous breaking of the PQ symmetry.
Its primary role is to solve the strong CP problem through the
dynamical relaxation of the QCD $\theta$ parameter.
The axion potential is generated by non-perturbative QCD instanton
effects, from which both the axion mass and its self-interaction are
uniquely determined.
In our previous work \cite{FukuSO10}, we have shown that this QCD
axion can naturally be identified with the relative phase along the
D-flat direction in the minimal supersymmetric SO(10) grand unified
theory.
For completeness, we briefly review the basic properties of the QCD
axion before introducing the string-inspired ALPs in
the next subsection.
QCD instanton potential in the classic dilute gas approximation becomes
\be
V(\phi)=m_u\Lambda_{QCD}^3\left[1-\cos\left(\frac{\phi}{f_a}\right)\right] \;.
\label{potential}
\ee
Using the Gellmann-Oakes-Renner relation \cite{Gellmann}, 
\be
\Lambda_{QCD}^3=\frac{F_\pi^2m_\pi^2}{m_u+m_d} \;.
\label{LQCD}
\ee
Here $F_\pi=93$ MeV, and $\frac{m_u}{m_d}\approx 0.47$, and we obtain axion mass
\be
m_a=5.7\times 10^{-6}\left(\frac{10^{12}\mbox{GeV}}{f_a}\right) \text{eV} \;.
\label{ma}
\ee

The intermediate scale appearing in Eq.~(\ref{ma}) has a natural
interpretation in the minimal supersymmetric SO(10) framework.
The symmetry-breaking chain from SO(10) to the Standard Model
through the intermediate Pati--Salam symmetry,
\be
SO(10)
\longrightarrow
SU(4)_C\times SU(2)_L\times SU(2)_R
\longrightarrow
SU(3)_C\times SU(2)_L\times U(1)_Y ,
\ee
was explicitly studied in Ref.~\cite{Fuku4}.
The breaking of the Pati--Salam symmetry is associated with vacuum
expectation values of the SM-singlet components contained in the
${\bf 126}$ and $\overline{\bf 126}$ Higgs multiplets.  Writing
schematically
\be
\langle\Delta_R\rangle
 = {v_R\over\sqrt{2}}e^{i\theta_1},
\qquad
\langle\overline{\Delta}_R\rangle
 = {\bar v_R\over\sqrt{2}}e^{i\theta_2},
\ee
supersymmetry requires both $F$- and $D$-flatness.  In particular,
the $D$-flatness condition implies
\be
|v_R|=|\bar v_R| .
\ee
This condition fixes the relative magnitudes of the two vacuum
expectation values but does not by itself fix their relative phase.
After removing the phase combination associated with the broken
gauge symmetry, a physical relative phase may remain.  If this
surviving degree of freedom is identified with the Peccei--Quinn
axion \cite{FukuSO10}, its canonically normalized field is schematically
\be
a\sim v_R(\theta_1-\theta_2),
\ee
up to normalization factors determined by the PQ charges and the
field content.  The same vacuum expectation values therefore set
both the Pati--Salam breaking scale and the PQ-breaking scale,
leading naturally to
\be
f_a=O(v_R)=O(M_{\rm PS}) .
\label{PSlevel}
\ee
For $v_R\sim10^{12}\ {\rm GeV}$ \cite{Fuku4,Fuku3}, this gives
$f_a\sim10^{12}\ {\rm GeV}$ and hence the QCD-axion mass scale
$m_a\sim10^{-6}\ {\rm eV}$ in Eq.~(\ref{ma}).
\subsection{Heterotic-string-inspired axion-like particles}

While the QCD axion provides an elegant solution to the strong CP
problem, modern superstring theory suggests a much richer axion
sector.
In particular, compactification of the heterotic string naturally
gives rise to a large number of ALPs, including
both the model-independent axion and model-dependent axions
associated with the topology of the internal compact manifold
\cite{Svrcek,Axiverse,Marsh}.
Unlike the QCD axion, whose mass is fixed by QCD instanton effects,
the masses of these string-inspired ALPs are generated by
non-perturbative effects and can therefore span an extremely wide
range.
This ``axiverse'' picture provides a natural theoretical framework in
which ultralight axions relevant to cosmology can arise.
Unlike the QCD axion discussed in the preceding subsection, the
decay constants of string-inspired axions are determined by the
normalization of the higher-dimensional antisymmetric tensor field
and its compactification to four dimensions.  In weakly coupled
heterotic string theory this normalization relates the characteristic
microscopic axion decay constant to the four-dimensional gauge
coupling and the reduced Planck scale. In the convention adopted
here, one obtains \cite{Svrcek}
\be
 f_a =
 \frac{\alpha_{\rm YM} M_{\rm Pl}^*}
 {2\sqrt{2}\pi}
 \simeq 1.1\times10^{16}\ {\rm GeV},
\label{heteroticfa}
\ee
with the reduced Planck mass
$M_{\rm Pl}^*=2.4\times10^{18}$ GeV.
This value provides a representative microscopic scale
for the heterotic axion sector.
The decay constants of individual model-dependent axions
depend on the geometry of the compactification,
as discussed in Sec.~4.

The instanton-induced potential is given by

\begin{equation}
V_i
=
\Lambda_i^4
\left[
1-
\cos
\left(
\frac{a_i}{f_i}
+\delta_i
\right)
\right].
\end{equation}

\begin{equation}
\Lambda_i^4
=
M_{\rm SUSY}^2
M_{\rm Pl}^{*2}
e^{-S_i}.
\label{Lambda4}
\end{equation}
Here $M_{\rm SUSY}$ parametrizes the supersymmetry-breaking
contribution entering the non-perturbative axion potential.
It should be distinguished from the SUSY threshold scale commonly
used in the renormalization-group analysis of supersymmetric grand
unified theories.  The latter denotes the characteristic mass scale
of the observable-sector superpartners, at which the running changes
from that of the Standard Model to that of its supersymmetric
extension.  By contrast, $M_{\rm SUSY}$ in Eq.~(\ref{Lambda4}) characterizes
the supersymmetry-breaking contribution to the prefactor of the
non-perturbative axion potential and need not coincide with the
low-energy superpartner mass scale.  In the present analysis we
therefore regard $M_{\rm SUSY}$ as a phenomenological parameter of
the heterotic-string-inspired axion sector, following
Ref.~\cite{Visinelli}.
In this parametrization, the axion mass spectrum is controlled by
the supersymmetry-breaking scale $M_{\rm SUSY}$ and by the instanton
actions $S_i$.  We treat $M_{\rm SUSY}$ as a common mass scale of the
heterotic-string-inspired axion sector, while the large hierarchy
among the axion masses is generated predominantly by the exponential
dependence on the different instanton actions $S_i$.
It is useful to estimate how different the instanton actions must be
in order to generate the two ultralight mass scales relevant to the
cosmological discussion below. 
From Eqs.~(\ref{heteroticfa})--(\ref{Lambda4}), the mass associated with a given
non-perturbative sector scales parametrically as
\be
 m_i \simeq
 \frac{M_{\rm SUSY}M_{\rm Pl}^*}{f_i}
 e^{-S_i/2}.
\label{massinstanton}
\ee
\textcolor{red}{
Equation~(\ref{massinstanton}) illustrates how an exponentially broad spectrum of
microscopic axion masses can arise from non-perturbative dynamics.
In a system containing several axions and several non-perturbative
sectors, however, the cosmologically relevant fields need not
coincide with the individual microscopic axions $a_i$.  Rather, they
are determined by the complete multidimensional potential and can
correspond to particular linear combinations of the microscopic
fields.
This distinction is important for the two ultralight cosmological
directions considered in Sec.~3.  The $10^{-22}\,{\rm eV}$
dark-matter direction may be treated as an ordinary light massive
eigenmode with a nonvanishing quadratic curvature.  By contrast, the
$10^{-28}\,{\rm eV}$ scale introduced in Sec.~3.3 characterizes the
cosmological dynamics of an aligned collective direction of the
multi-axion--multi-instanton potential.  It should therefore not be
identified directly with the microscopic mass $m_i$ in Eq.~(\ref{massinstanton}).
Consequently, although Eq.~(\ref{massinstanton}) demonstrates that the microscopic
axion sector can naturally contain exponentially separated mass
scales, we do not infer the hierarchy between the two cosmologically
relevant scales from the ratio of two individual instanton-induced
masses.  Their different physical origins and their relation to the
underlying multi-axion potential will be discussed explicitly in
Sec.~3.3.
}



\section{Cosmological implications of ultralight axion-like particles}
\textcolor{red}{
Having discussed the theoretical origin of the QCD axion and
string-inspired ALPs, we now turn to their cosmological implications.
The important point is that the axiverse by itself does not determine
which directions in the multidimensional axion field space are
cosmologically relevant.  Conversely, cosmological phenomena can
provide selection criteria for particular directions and for the
structure of their effective potentials within the much larger
multi-axion sector.
In the present framework, the QCD axion and the two cosmologically
relevant ultralight directions play qualitatively different roles.
The conventional QCD axion, with a mass of order
\begin{equation}
 m_a=O(10^{-6})\,{\rm eV},
\end{equation}
is associated primarily with the solution of the strong CP problem.
The first cosmologically relevant ultralight direction is a light
massive eigenmode with
\begin{equation}
 m_{22}=O(10^{-22})\,{\rm eV}.
\end{equation}
It is assumed to constitute the dominant dark-matter component and
provides the characteristic scale relevant to Bose--Einstein
condensation and the subsequent nonlinear collapse considered in
Secs.~3.1 and 3.2.
The second is qualitatively different.  It is an aligned collective
direction of the multi-axion field space with a characteristic
dynamical scale
\begin{equation}
 m_{28}^{\rm char}=O(10^{-28})\,{\rm eV}.
\end{equation}
This scale is chosen so that the evolution of the collective field
becomes important near the matter--radiation transition, where a
transient contribution to the energy density can modify the
pre-recombination expansion history.  As emphasized in Sec.~2.2,
$m_{28}^{\rm char}$ is not identified with the small-oscillation mass
of an individual microscopic axion.
Thus the two ultralight scales should not be viewed simply as two
independently chosen ALP masses.  They characterize two
qualitatively different directions of the same underlying
multi-axion sector: a massive dark-matter eigenmode with nonvanishing
quadratic curvature and an aligned collective direction whose
lower-order curvature can be strongly suppressed by the projected
multi-instanton contributions.
In the following, we first discuss BEC formation and nonlinear
collapse of the $10^{-22}\,{\rm eV}$ dark-matter mode and its
possible role in the early formation of supermassive black holes.
We then consider the collective $10^{-28}\,{\rm eV}$ direction and
its possible role in modifying the pre-recombination expansion
history relevant to the Hubble tension.
}

\subsection{BEC formation}

We first consider the cosmological evolution of the ultralight
axion-like particle with a characteristic mass, $m_a \sim 10^{-22}\ {\rm eV}$.
Because of its extremely small mass, the corresponding de Broglie
wavelength can become astrophysically large, while the occupation
number of the low-momentum modes can be enormous.  The axion field
may therefore behave as a highly coherent bosonic system rather
than as a dilute gas of individual particles.  Under appropriate
cosmological conditions, such a system can undergo BEC, leading to a macroscopic occupation of the
lowest-energy states \cite{FMT}.

The formation and subsequent evolution of the condensate are
particularly important when the axion self-interaction is
attractive.  In this case, the competition among gradient pressure,
self-interaction, and gravity can drive the condensate away from a
stationary configuration and eventually trigger a nonlinear
collapse.  The resulting dynamics is not necessarily a single
collapse event: after a sufficiently dense configuration is formed,
particle emission and redistribution can reduce the central density,
allowing the system to recondense and collapse again.  The
cosmological axion condensate can therefore exhibit a sequence of
condensation, collapse, and recondensation.

This nonlinear BEC dynamics provides the starting point of our
cosmological scenario.  In the following we first summarize the
conditions for axion BEC formation and its characteristic scales,
and then discuss how the resulting collapse dynamics can provide
massive seeds for the early formation of supermassive black holes.

\subsection{Early supermassive black holes}
\label{sec:SMBH}

We now turn to the possible role of the
$m_a=O(10^{-22})\,{\rm eV}$ axion component in the early formation
of supermassive black holes.  In the present framework this axion
is assumed to constitute the dominant dark-matter component.
Because of its extremely small mass, it can develop a macroscopically
coherent state and, under appropriate conditions, form a
BEC.  As discussed in the preceding
subsection, an attractive self-interaction can then render the
condensate unstable and drive it toward nonlinear collapse.

The possible relevance of such an ultralight axion to the
high-redshift SMBH problem was discussed in
Refs.~\cite{FukuyamaHT, FukuyamaSMBH}.  The basic difficulty is that
very massive black holes are already observed at high redshift,
although ordinary baryonic accretion and the hierarchical growth of
collisionless dark matter require substantial cosmic time.  The
problem becomes even more severe if the large angular momentum of
the collapsing system is taken into account.  In
Ref.~\cite{FukuyamaSMBH} it was pointed out that an attractive axion
self-interaction can provide an additional mechanism for overcoming
the centrifugal barrier and thereby accelerate the formation of a
compact object.

To see the origin of this effect, consider the nonrelativistic
effective potential of an axion condensate.  Near the minimum of an
ordinary axion potential one may write schematically
\be
 V_{22}(\phi_{22})
 =
 \frac{1}{2}m_{22}^2\phi_{22}^2
 +
 \frac{\lambda_{22}}{4}\phi_{22}^4+\cdots ,
\label{V22}
\ee
with
\be
 \lambda_{22}<0
\label{lambda22}
\ee
for an attractive self-interaction.  For the ordinary cosine
potential,
\be
 V_{22}
 =
 m_{22}^2f_{22}^2
 \left[
 1-\cos\left(\frac{\phi_{22}}{f_{22}}\right)
 \right],
\ee
the expansion around the minimum gives
\be
 V_{22}
 =
 \frac{1}{2}m_{22}^2\phi_{22}^2
 -
 \frac{m_{22}^2}{24f_{22}^2}\phi_{22}^4
 +O(\phi_{22}^6).
\label{cosine22}
\ee
Comparing this expansion with Eq.~(\ref{V22}),
we obtain
\be
 \lambda_{22}
 =
 -\frac{m_{22}^2}{6f_{22}^2}<0,
\ee
so that the quartic interaction is indeed attractive.

In a self-gravitating condensate the competition among quantum
pressure, gravity, angular momentum, and the attractive
self-interaction determines the stability of the configuration.
Schematically, for a condensate characterized by a size $R$ and
particle number $N$, the energy contains contributions of the form
\be
 E(R)
 \sim
 \frac{N}{2m_{22}R^2}
 +
 \frac{L^2}{2m_{22}NR^2}
 -
 \frac{Gm_{22}^2N^2}{R}
 +
 \frac{\lambda_{22}N^2}{m_{22}^2R^3}.
\label{energySMBH}
\ee
The first term represents quantum pressure, the second the
centrifugal contribution, the third gravitational attraction, and
the last the self-interaction.  Since $\lambda_{22}<0$, the last
term lowers the energy as the condensate contracts and can remove
the metastable minimum which would otherwise be supported by
gradient and rotational effects.  Once the critical configuration
is crossed, the system undergoes a rapid collapse.

This observation is particularly relevant to the apparent tension
between the early formation of SMBHs and their large angular
momenta.  The attractive interaction does not merely supplement
gravity; because its contribution grows more rapidly as the size
decreases,
\be
 |E_{\rm self}|\propto R^{-3},
\qquad
 |E_{\rm grav}|\propto R^{-1},
\ee
it becomes increasingly important during contraction.  It can
therefore trigger a nonlinear instability even when the
centrifugal barrier would inhibit purely gravitational collapse
\cite{FukuyamaSMBH,FukuyamaHT}.

The characteristic ultralight mass scale is also important.
An axion with
\be
 m_{22}=O(10^{-22})\,{\rm eV}
\label{m22SMBH}
\ee
has an astrophysically large de Broglie wavelength and can behave
coherently on galactic or subgalactic scales.  Such masses have long
been considered in fuzzy- or ultralight-dark-matter scenarios
\cite{HuBarkanaGruzinov,HuiEtAl,DiezTejedorMarsh}.  In particular,
ultralight axions in this mass range can constitute a substantial,
and in appropriate cosmological histories essentially the entire,
dark-matter abundance \cite{DiezTejedorMarsh}.

The relation between condensation and collapse was investigated
further in Ref.~\cite{FukuyamaBEC}.  There it was found that the
formation of the axion BEC can occur at an earlier epoch,
characteristically around
\be
 z_{\rm BEC}=O(30),
\ee
with the condensation rate being dominated primarily by
gravitational interactions, whereas the comparatively weak
attractive self-interaction is crucial for destabilizing the
condensate after it has formed.  The two interactions therefore
play complementary roles:
\be
 {\rm gravity}
 \quad\longrightarrow\quad
 {\rm condensation},
\qquad
 {\rm attractive\ self\!-\!interaction}
 \quad\longrightarrow\quad
 {\rm instability\ and\ collapse}.
\label{gravityself}
\ee
This separation of roles is important: the self-interaction need
not be large enough to generate the condensate itself in order to
have a decisive effect on its subsequent nonlinear evolution.
\textcolor{red}{
We next clarify the initial condition and the time scale associated
with the nonlinear collapse.  The mechanism considered here does not
require an anomalously large primordial density perturbation generated
specifically, for example, by post-inflationary PQ breaking.  The
general-relativistic collapse analysis of Ref.~\cite{FukuyamaCollapse}, however, starts
from an overdense axion region which has already become sufficiently
nonlinear to decouple from the cosmological expansion.  Thus an
overdense region is part of the initial condition of the collapse
problem, but the essential mechanism driving the final instability is
the attractive axion self-interaction rather than an assumed
exceptionally large primordial fluctuation.
The collapse time and length scales were studied explicitly in
Ref.~\cite{FukuyamaCollapse}.  Near the critical configuration the self-attracting
condensate becomes approximately pressureless, so that its subsequent
collapse can be described, to a useful approximation, by a
spherically symmetric dust configuration in general relativity.  The
characteristic collapse time obtained in this treatment is
\begin{equation}
 \tau_c =
 \frac{\pi}{2} r_g
 \left[
 \left(\frac{R}{r_g}\right)^2+1
 \right]^{3/2},
\end{equation}
where $R$ is the initial radius of the collapsing configuration and
$r_g$ denotes its gravitational radius.  The time required for the
configuration to reach the strongly collapsing regime is
approximately
\begin{equation}
 t_{\rm col}\simeq \frac{\tau_c}{2}.
\end{equation}
For the representative critical configuration considered in
Ref.~\cite{FukuyamaCollapse}, with
\begin{equation}
 R_0\simeq 2.3~{\rm pc},
 \qquad
 M\simeq 5\times10^9\,M_\odot ,
\end{equation}
one obtains the characteristic estimate
\begin{equation}
 t_{\rm col}=O(10^8)~{\rm yr}.
\end{equation}
This is short enough to be compatible with the appearance of massive
compact objects at high redshift once the condensate has reached the
critical configuration.
It is important, however, to distinguish this dynamical collapse time
from the cosmological time at which a given axion overdensity first
reaches the instability threshold.  The latter depends on the initial
overdensity and on the preceding nonlinear evolution of the axion
condensate and is not determined by the axion mass alone.  The result
$t_{\rm col}=O(10^8)\,{\rm yr}$ should therefore be understood as the
collapse time after the critical self-attracting configuration has
formed, rather than as a prediction of the redshift at which collapse
necessarily begins.
The resulting compact object need not immediately possess the final
mass of an observed SMBH.  Rather, the axion collapse may provide an
early massive seed whose subsequent evolution through accretion and
mergers can lead to the SMBHs observed at $z\gtrsim7$.
The cosmological picture can therefore be summarized as
\be
\begin{aligned}
{\rm ultralight\ axion\ dark\ matter}
&\longrightarrow {\rm BEC},
\\
{\rm BEC}
&\longrightarrow {\rm self\!-\!interaction\ instability}
\longrightarrow {\rm nonlinear\ collapse},
\\
{\rm nonlinear\ collapse}
&\longrightarrow {\rm massive\ seed}.
\end{aligned}
\label{SMBHsequence}
\ee
This mechanism also explains why the $10^{-22}$ eV axion plays a
different role from the $10^{-28}$ eV collective direction discussed
in Sec.~3.3. The former constitutes the dominant dark matter and
undergoes nonlinear, spatially inhomogeneous condensate dynamics,
whereas the latter is introduced primarily to modify the homogeneous
pre-recombination expansion history. The two directions therefore
probe distinct aspects of the broad axion field space.}
Finally, repeated collapse and recondensation of the
$10^{-22}$ eV condensate may leave additional signatures in the
high-redshift Universe.  In particular, a sequence of recurrent
collapse events has been suggested as a possible origin of the
approximately log-periodic modulation reported in the abundance
of high-redshift compact objects
\cite{FukuyamaLogPeriodic}.  If the
collapse of the $m_a=O(10^{-22})\,{\rm eV}$ condensate is followed
by redistribution and recondensation, the system may undergo a
sequence of recurrent collapse events.  When successive events are
approximately self-similar, their characteristic scale factors may
satisfy
\be
 \frac{a_{n+1}}{a_n}\simeq q ,
\ee
with an approximately constant ratio $q$.  Since
$a=(1+z)^{-1}$, this corresponds to an approximately constant
spacing in
\be
 x=\ln(1+z).
\ee
The resulting high-redshift abundance may then contain a modulation
of the form
\be
 {\cal N}(z)
 =
 {\cal N}_0(z)
 \left[
 1+A\cos\left\{
 \omega\ln(1+z)+\varphi
 \right\}
 \right],
\ee
with
\be
 \Delta\ln(1+z)=\frac{2\pi}{\omega}.
\ee
Evidence for such a modulation in the abundance of high-redshift
compact objects was reported in
Ref.~\cite{FukuyamaLogPeriodic}.  We do not develop this aspect
further here; rather, we regard it as a possible additional
observational consequence of the same repeated
collapse--recondensation dynamics responsible for the early
formation of massive compact objects.
\subsection{Hubble tension}
\label{sec:Hubble}
We next consider the possible cosmological role of an aligned
collective direction with a characteristic dynamical scale
\begin{equation}
m_{28}^{\rm char}=O(10^{-28})~{\rm eV}.
\label{m28}
\end{equation}
This component should be distinguished from the
$m_a=O(10^{-22})$ eV axion discussed in the preceding subsections,
which is assumed to constitute the dominant dark-matter component
and whose attractive self-interaction may be relevant to the early
formation of supermassive black holes. The collective direction
considered here instead affects the homogeneous expansion history
around the epoch of matter--radiation equality.

Ultralight axion-like fields remain approximately frozen by Hubble
friction when the expansion rate is sufficiently large compared
with their characteristic dynamical scale
\cite{Bella, PoulinULA, PoulinEDE}.
The homogeneous field equation is
\be
 \ddot{\phi}_{28}+3H\dot{\phi}_{28}
 +\frac{\partial V_{28}}{\partial\phi_{28}}=0.
\label{KG28}
\ee
For the potential constructed in Sec.~4,
\be
 V_{28}(\phi_{28})=
 \Lambda_{28}^4
 \left(1-\cos\frac{\phi_{28}}{f_{\rm eff}}\right)^3,
 \label{V28cube}
\ee
the curvature at the minimum vanishes,
$V_{28}''(0)=0$.
We therefore introduce the characteristic dynamical scale
\be
 m_{28}^{\rm char}\equiv
 \frac{\Lambda_{28}^2}{f_{\rm eff}},
\label{m28char}
\ee
rather than the conventional axion mass defined by the
curvature of the potential at its minimum.

As an order-of-magnitude characterization, the onset of the
evolution may be written as
\be
 H(z_c)\sim m_{28}^{\rm char},
\label{zc}
\ee
where $z_c$ denotes the critical redshift at which the ultralight
axion field begins to evolve dynamically.
The precise onset depends on the initial displacement and
the shape of the potential.
For comparison, denoting the redshift of matter--radiation
equality by $z_{\rm eq}$, the Hubble scale at this epoch is
of order
\be
 H(z_{\rm eq})\approx 2.3\times10^{-28}\ {\rm eV}.
\label{zeq}
\ee
Thus, the characteristic dynamical scale
$m_{28}^{\rm char}=O(10^{-28})~{\rm eV}$
places the onset of the axion dynamics close to the epoch
at which a modification of the pre-recombination expansion
history can affect the acoustic scale.
The string axiverse provides a natural framework for axion
scales extending over many orders of magnitude
\cite{Axiverse}.

 In contrast to the nonlinear and spatially inhomogeneous dynamics
discussed in Sec.~3.2, the present mechanism concerns the homogeneous
pre-recombination expansion history.  It also differs from the
treatment adopted in Ref.~\cite{FukuyamaHT}. The comoving sound horizon at photon
decoupling is always defined by
\be
 r_s(z_*)=
 \int_{z_*}^{\infty}
 \frac{c_{s,\gamma b}(z)}{H(z)}\,dz,
\label{rs_correct}
\ee
where
\be
 c_{s,\gamma b}^2=
 \frac{1}{3(1+R)},\qquad
 R=\frac{3\rho_b}{4\rho_\gamma}.
\label{cs_photon}
\ee
The upper limit of this integral is not determined by the onset of
axion oscillations.  In particular, the freezing of the axion field
does not imply the disappearance of acoustic propagation in the
photon--baryon plasma.  The axion contribution must instead modify
the sound horizon through the expansion rate $H(z)$.

We therefore write
\be
 3M_{\rm Pl}^{*\,2}H^2
 =
 \rho_r+\rho_b+\rho_{\rm DM}
 +\rho_{28}+\rho_\Lambda ,
\label{Friedmann28}
\ee
where the dominant dark-matter density includes the
$m_a=O(10^{-22})~{\rm eV}$ axion component, while
\be
 \rho_{28}
 =
 \frac{1}{2}\dot{\phi}_{28}^{\,2}+V_{28}(\phi_{28}).
\label{rho28}
\ee
A positive contribution $\rho_{28}$ around the pre-recombination
epoch increases $H(z)$ and consequently reduces the sound horizon,
\be
 H(z)>H_{\Lambda{\rm CDM}}(z)
 \quad\Longrightarrow\quad
 r_s(z_*)<r_s^{\Lambda{\rm CDM}}(z_*).
\label{rsreduce}
\ee
Such a transient modification of the pre-recombination expansion
rate is the basic mechanism underlying early-dark-energy approaches
to the Hubble tension \cite{KarwalKamionkowski,PoulinEDE}.

Since the acoustic angular scale
\be
 \theta_*=\frac{r_s(z_*)}{D_M(z_*)}
\label{thetaac}
\ee
is accurately determined by the CMB, a reduction of the physical
sound horizon allows a larger inferred value of $H_0$ while
preserving the observed acoustic scale.

\subsubsection{Microscopic and effective axion scales}

An important point arises when this mechanism is embedded in the
heterotic axion sector discussed in Sec.~2.  For an individual
microscopic heterotic axion, the decay constant is naturally close
to the GUT scale (Eq.~(\ref{heteroticfa})) \cite{Svrcek, Axiverse} .
We retain this relation as the microscopic axion scale.

The cosmologically relevant field, however, need not coincide with
one microscopic axion.  In a theory containing several axions, the
nonperturbative potential can take the general form
\be
 V(\{a_i\})
 =
 \sum_A\Lambda_A^4
 \left[
 1-\cos\left(
 \sum_i Q_{Ai}\frac{a_i}{f_i}+\delta_A
 \right)
 \right].
\label{multiV}
\ee
The mass eigenstates are then linear combinations of the microscopic
axion fields.  Alignment among the instanton charge vectors can
generate a light direction whose effective decay constant is much larger
than the individual microscopic decay constants \cite{KNP,ChoiKimYun,HigakiTakahashi}.
We write the canonically normalized light direction as
\be
 \phi_{28}=\sum_i c_i a_i,
 \qquad
 \sum_i c_i^2=1 .
\label{lightdirection}
\ee
Along this direction, $a_i=c_i\phi_{28}$, and the phase associated
with a nonperturbative sector $A$ becomes
\be
 \sum_i Q_{Ai}\frac{a_i}{f_i}
 =
 \phi_{28}\sum_i\frac{Q_{Ai}c_i}{f_i}.
\ee

If the projected charges along the light axion direction
are commensurate, the effective potential is periodic
with a common period $2\pi f_{\rm eff}$.
We may then write
\be
 \sum_i\frac{Q_{Ai}c_i}{f_i}
 =
 \frac{k_A}{f_{\rm eff}},
\label{feff}
\ee
where the integer $k_A$ labels the harmonic associated
with the $A$th non-perturbative sector.

For an aligned light direction the projected charge can be small,
allowing
\be
 f_{\rm eff}\gg f_i ,
\label{alignment}
\ee
where $f_{\rm eff}$ denotes the effective decay constant of the
light axion direction.  After the heavier axionic directions are
stabilized, the low-energy potential may then be written schematically
as
\be
 V_{28}(\phi_{28})
 =
 \Lambda_{28}^4
 {\cal U}\left(\frac{\phi_{28}}{f_{\rm eff}}\right).
\label{V28}
\ee
For the potential adopted in Eq.~(\ref{V28cube}),
\be
 U(\theta)=(1-\cos\theta)^3,
\ee
the energy density before the onset of the evolution
is approximately
\be
 \rho_{28}\simeq
 (m_{28}^{\rm char})^2 f_{\rm eff}^2
 (1-\cos\theta_{\rm ini})^3,
 \qquad
 \theta_{\rm ini}\equiv
 \frac{\phi_{28,\rm ini}}{f_{\rm eff}},
 \label{rhoinitial}
\ee
where $\phi_{28,\rm ini}$ denotes the initial displacement
of the effective ultralight mode.
Introducing $\kappa_c\equiv H(z_c)/m_{28}^{\rm char}$,
one has
\be
 \rho_{\rm tot}(z_c)=
 3(M_{\rm Pl}^*)^2\kappa_c^2
 (m_{28}^{\rm char})^2,
\ee
and hence, neglecting the kinetic energy and the
change in the field displacement before $z_c$,
\be
 \mathcal{F}_{28}(z_c)
 \equiv\frac{\rho_{28}(z_c)}{\rho_{\rm tot}(z_c)}
 \simeq
 \frac{f_{\rm eff}^2}
 {3\kappa_c^2(M_{\rm Pl}^*)^2}
 (1-\cos\theta_{\rm ini})^3.
\ee
For fixed $\theta_{\rm ini}$ and $\kappa_c$,
the characteristic dynamical scale primarily determines
the epoch at which the field becomes dynamical,
whereas the effective decay constant controls
the magnitude of its contribution:
\be
 m_{28}^{\rm char}\longrightarrow z_c,
 \qquad
 f_{\rm eff}\longrightarrow\mathcal{F}_{28}(z_c).
\ee

The distinction between the microscopic decay constants $f_i$ and
the effective decay constant $f_{\rm eff}$ is directly relevant to the
cosmological role of the $10^{-28}$ eV mode.  While the microscopic
heterotic scales remain of order $10^{16}$ GeV, the aligned light
direction can have a substantially larger effective decay constant and
can therefore store a non-negligible fraction of the total energy
density around the onset of its cosmological evolution.  The
instantaneous fraction $\mathcal{F}_{28}(z_c)$ estimated above does not,
however, directly determine the change in the sound horizon, which
depends on the evolution of this component over the entire
pre-recombination epoch.  We therefore next evaluate the
sound-horizon-weighted contribution of the light mode and quantify
the resulting reduction of $r_s$.
\subsubsection{Modification of the sound horizon}

We now examine how the light axion mode modifies the sound horizon.
Defining
\be
 \mathcal{F}_{28}(z)
 =
 \frac{\rho_{28}(z)}
 {\rho_{\rm std}(z)+\rho_{28}(z)},
\label{F28}
\ee
the Friedmann equation may be written as
\be
 H(z)
 =
 \frac{H_{\rm std}(z)}
 {\sqrt{1-\mathcal{F}_{28}(z)}} .
\label{HF28}
\ee
Substituting this expression into Eq.~(\ref{rs_correct}), we obtain
\be
 r_s(z_*)
 =
 \int_{z_*}^{\infty}
 \frac{c_{s,\gamma b}(z)}
      {H_{\rm std}(z)}
 \sqrt{1-\mathcal{F}_{28}(z)}\,dz .
\label{rsF28}
\ee
For a sufficiently small contribution,
\be
 \frac{\Delta r_s}{r_s}
 \simeq
 -\frac{1}{2}
 \left\langle \mathcal{F}_{28}\right\rangle_s ,
\label{drs}
\ee
where
\be
 \left\langle \mathcal{F}_{28}\right\rangle_s
 =
 \frac{
 \displaystyle
 \int_{z_*}^{\infty} dz\,
 \frac{c_{s,\gamma b}(z)}
      {H_{\rm std}(z)}
 \mathcal{F}_{28}(z)}
 {
 \displaystyle
 \int_{z_*}^{\infty} dz\,
 \frac{c_{s,\gamma b}(z)}
      {H_{\rm std}(z)}
 }
\label{weightedF}
\ee
is the sound-horizon-weighted fractional contribution of the light
axion component.
To estimate the magnitude of the axion contribution required by the
Hubble tension, we note that the accurately measured CMB acoustic
scale implies, to a useful first approximation,
\be
H_0 r_s \simeq {\rm const.}
\label{H0rs}
\ee
when the modification is confined mainly to the pre-recombination
expansion history.  Taking representative values
$H_0^{\rm CMB}\simeq67.4~{\rm km\,s^{-1}\,Mpc^{-1}}$ and
$H_0^{\rm local}\simeq73~{\rm km\,s^{-1}\,Mpc^{-1}}$, the sound
horizon would have to be reduced approximately by
\be
\frac{r_s^{\rm new}}{r_s^{\Lambda{\rm CDM}}}
\simeq
\frac{67.4}{73}
\simeq0.923 ,
\ee
or
\be
\frac{\Delta r_s}{r_s}\simeq-0.077.
\label{requiredrs}
\ee
Using Eq.~(\ref{drs}), this corresponds to a sound-horizon-weighted
axion fraction
\be
\left\langle \mathcal{F}_{28}\right\rangle_s
\simeq0.15 .
\label{requiredf28}
\ee
The peak axion fraction must be at least as
large as the sound-horizon-weighted fraction,
\begin{equation}
\mathcal{F}_{28}^{\rm peak}
\geq
\langle\mathcal{F}_{28}\rangle_s
\simeq0.15.
\end{equation}
The precise relation between these quantities
requires a numerical evolution of the axion
field and the cosmological background.
This estimate also gives a useful indication of the effective field
range required by the multi-axion sector.  
Using the sixth-order potential and taking a large
initial displacement, $\theta_{\rm ini}\simeq\pi$,
one obtains
\be
 \mathcal{F}_{28}(z_c)\simeq
 \frac{8f_{\rm eff}^2}
 {3\kappa_c^2(M_{\rm Pl}^*)^2}.
\ee
For an illustrative value
$\mathcal{F}_{28}(z_c)\sim0.15$, this gives
\be
 f_{\rm eff}\simeq
 \kappa_c M_{\rm Pl}^*
 \sqrt{\frac{3\mathcal{F}_{28}(z_c)}{8}}
 \simeq
 5.7\times10^{17}\kappa_c~{\rm GeV}.
\ee
For $\kappa_c=O(1)$, the effective decay constant
is therefore of order $10^{18}$ GeV.
Compared with the representative microscopic
heterotic scale, the required enhancement is
approximately
\be
 \frac{f_{\rm eff}}{f_i}=O(10^2).
\ee
This provides an order-of-magnitude target for
the multi-axion alignment required in the present
scenario.

Equations~(\ref{rsF28}) and (\ref{drs}) replace the truncation of
the sound-horizon integral employed in our previous analysis
\cite{FukuyamaHT}.  The reduction of $r_s$ is now generated
dynamically by the axion contribution to the Friedmann equation,
rather than by terminating acoustic propagation at the onset of
axion evolution.  A transient increase of the pre-recombination
expansion rate is the basic mechanism employed in early-dark-energy
approaches to the Hubble tension
\cite{KarwalKamionkowski,SPA, PoulinEDE}.

A further requirement is that the additional axion component be
transient.  It should become cosmologically relevant around the
critical epoch $z_c$, increase $H(z)$ and thereby reduce $r_s$, but
its fractional contribution should subsequently decrease rapidly
enough that the standard late-time cosmological evolution is
recovered.
For an isolated axion generated by a single dominant instanton
sector, the potential is approximately of the ordinary cosine form,
\be
 V(\phi)
 =
 \Lambda^4
 \left[
 1-\cos\left(\frac{\phi}{f}\right)
 \right].
\label{singlecos}
\ee
Near its minimum this potential is quadratic.  Consequently, after
the onset of coherent oscillations the field behaves on average as
a pressureless matter component,
\be
 \langle w\rangle\simeq0,
 \qquad
 \rho_\phi\propto a^{-3}.
\label{cosmatter}
\ee
A single-cosine axion therefore does not naturally provide the rapid
post-critical dilution desired here
\cite{PoulinULA}.
The situation is qualitatively different in the multi-axion
framework of Eq.~(\ref{multiV}).  It is useful to emphasize that the
different instanton sectors should not be regarded as independently
producing axions with masses that are subsequently superposed.
Rather, the complete non-perturbative potential determines an axion
mass matrix,
\be
 (M_a^2)_{ij}
 =
 \left.
 \frac{\partial^2 V}
      {\partial a_i\partial a_j}
 \right|_{\rm min}
 =
 \sum_A
 \frac{\Lambda_A^4 Q_{Ai}Q_{Aj}}
      {f_i f_j}
 \cos\Theta_A\bigg|_{\rm min},
\label{axionmassmatrix}
\ee
where
\be
 \Theta_A
 =
 \sum_i Q_{Ai}\frac{a_i}{f_i}+\delta_A .
\ee
\textcolor{red}{
Diagonalization of the quadratic mass matrix determines the ordinary
mass eigen-directions near a minimum.  The $10^{-28}$ eV component
considered here, however, is qualitatively different.  It is an
aligned collective direction in the full multi-axion field space
along which the lower-order curvature can be strongly suppressed by
the combined non-perturbative contributions.  We denote this
collective direction by $\phi_{28}$, as defined in Eq.~(\ref{lightdirection}).  After
the orthogonal heavier axionic directions are stabilized, the
different non-perturbative sectors project onto this same collective
direction.  The resulting one-dimensional effective potential is
therefore generically multi-harmonic,
}
\be
 V_{\rm eff}(\theta)
 =
 \sum_k A_k
 \left[
 1-\cos(k\theta+\delta_k)
 \right],
 \qquad
 \theta
 =
 \frac{\phi_{28}}{f_{\rm eff}} .
\label{multiharmonic}
\ee
\textcolor{red}{
Here the different harmonics do not represent different axion
particles.  They are different non-perturbative contributions to
the potential along the same collective direction $\phi_{28}$.
Different field-space directions correspond to different projections
of the instanton charge vectors and therefore probe different
combinations of the non-perturbative sectors.  The $10^{-22}$ eV
dark-matter direction may thus retain a nonvanishing quadratic
curvature, whereas along the $10^{-28}$ eV direction the lower-order
curvatures can be suppressed by cancellations among the projected
contributions.
}

\textcolor{red}{
The connection with the higher-order potential used below can be
shown explicitly.  For $n=3$, the exact identity
}
\be
\textcolor{red}{
(1-\cos\theta)^3
=
\frac{5}{2}
-\frac{15}{4}\cos\theta
+\frac{3}{2}\cos 2\theta
-\frac{1}{4}\cos 3\theta
}
\label{harmonic}
\ee
\textcolor{red}{
shows that a three-harmonic potential with $k=1,2,3$ and suitable
relative amplitudes and phases reproduces
}
\be
\textcolor{red}{
V_{28}(\phi_{28})
=
\Lambda_{28}^4
\left[
1-\cos\left(
\frac{\phi_{28}}{f_{\rm eff}}
\right)
\right]^3 .
}
\label{V28n}
\ee
\textcolor{red}{
Thus this potential need not be introduced as an independent
single-field ansatz: it can arise as the effective potential along
a collective direction of the multi-axion--multi-instanton system.
Near the minimum,
}
\be
\textcolor{red}{
V_{28}(\phi_{28})
=
\frac{\Lambda_{28}^4}{8}
\left(
\frac{\phi_{28}}{f_{\rm eff}}
\right)^6
+O(\phi_{28}^8),
}
\label{V6}
\ee
\textcolor{red}{
so that
}
\be
\textcolor{red}{
V_{28}''(0)=0,\qquad
V_{28}^{(4)}(0)=0,\qquad
V_{28}^{(6)}(0)\neq0 .
}
\label{derivs}
\ee

\textcolor{red}{
More generally, a convenient parametrization of the effective
potential along such a flattened collective direction is
}
\be
 V_{28}(\phi_{28})
 =
 \Lambda_{28}^4
 \left[
 1-\cos\left(
 \frac{\phi_{28}}{f_{\rm eff}}
 \right)
 \right]^n .
\label{EDEpotential}
\ee
\textcolor{red}{
For $n>1$ the usual quadratic curvature at the minimum is absent,
and the leading behavior is
$V_{28}\propto\phi_{28}^{2n}$.  The $n=3$ case above provides an
explicit multi-harmonic realization of the potential relevant to the
transient cosmological component.  We emphasize that this demonstrates
the consistency of the effective potential with the
multi-axion--multi-instanton framework; it does not constitute a
derivation from a specific heterotic compactification.
}
For coherent oscillations in a potential
$V\propto |\phi|^{2n}$, provided that the oscillation
period is sufficiently short compared with the Hubble
time, the oscillation-averaged equation of state is
\cite{PoulinULA, SPA}
\be
 \langle w_{28}\rangle
 =
 \frac{n-1}{n+1},
\label{w28}
\ee
and hence
\be
 \rho_{28}
 \propto
 a^{-6n/(n+1)}.
\label{rho28dilution}
\ee
The requirement that the axion energy density
dilute faster than radiation implies
\be
 \frac{6n}{n+1}>4,
 \qquad n>2.
\ee
Thus $n=3$ is the smallest integer exponent
satisfying this requirement and gives
\be
 \rho_{28}\propto a^{-9/2},
\label{9/2}
\ee
which decreases faster than radiation.
Larger values of $n$ are not excluded by
this cosmological requirement alone.
As discussed above, for $n>1$ the curvature
of the potential at the minimum vanishes.
The relevant cosmological time scale is therefore
characterized by $m_{28}^{\rm char}$ defined in
Eq.~(\ref{m28char}).

For the potential in Eq.~(\ref{EDEpotential}),
the field-dependent curvature is
\be
 V_{28}''(\phi_{28})
 =
 (m_{28}^{\rm char})^2 n
 (1-\cos\theta)^{n-1}
 (n-1+n\cos\theta),
 \qquad
 \theta=\frac{\phi_{28}}{f_{\rm eff}}.
\ee
An approximate criterion for the onset of the
field evolution is
\be
 H^2(z_c)
 \sim
 \left|
 V_{28}''(\phi_{28,{\rm ini}})
 \right|.
\label{onsetgeneral}
\ee
The characteristic value
$m_{28}^{\rm char}=O(10^{-28})$ eV therefore
specifies the cosmological time scale of the
collective direction, while the detailed onset
redshift depends on the shape of the effective
potential and on the initial field displacement.

The picture that emerges is therefore the following.  A number of
microscopic heterotic axions with decay constants
$f_i=O(10^{16})$ GeV and several non-perturbative sectors generate
a multidimensional periodic potential.   Alignment can produce a collective direction with an enhanced
effective decay constant. Cosmology selects a characteristic
dynamical scale of order $10^{-28}$ eV for this direction, so that
it becomes active near the matter--radiation transition. The same
multi-instanton structure can in principle generate a
multi-harmonic potential along this direction; if the lower-order
terms are sufficiently suppressed, the resulting effective
potential can be represented by Eq.~(\ref{EDEpotential}) and can
dilute rapidly after the critical epoch.
\textcolor{red}{
An important consequence of this construction is that the
$10^{-22}$ eV and $10^{-28}$ eV sectors are not two completely
independent additions to cosmology.  They correspond to different
field-space directions of the same underlying
multi-axion--multi-instanton system and therefore depend on the same
microscopic set of decay constants, instanton charge vectors,
non-perturbative scales, and phases,
}
\be
\textcolor{red}{
 \{f_i,Q_{Ai},\Lambda_A,\delta_A\}.
}
\label{micro}
\ee
\textcolor{red}{
Cosmology nevertheless requires these two directions to have
qualitatively different local structures.  The dark-matter direction
$\phi_{22}$ must retain a nonvanishing quadratic curvature,
$V_{22}''(0)\neq0$, so that it behaves as an ultralight massive field
and can participate in the condensate dynamics discussed in
Sec.~3.2.  By contrast, the pre-recombination direction $\phi_{28}$
must be sufficiently aligned and flattened that its quadratic and
quartic curvatures are suppressed,
$V_{28}''(0)\simeq V_{28}^{(4)}(0)\simeq0$, while a higher-order term
remains.  At the same time, the sound-horizon requirement gives
$f_{\rm eff}=O(10^{18})$ GeV, approximately two orders of magnitude
above the characteristic microscopic heterotic scale
$f_i=O(10^{16})$ GeV.  Thus the two cosmological phenomena jointly
constrain the geometry of the same axion field space: one direction
must remain an ordinary light massive direction, whereas another
must exhibit both enhanced effective decay constant and suppressed
lower-order curvature.  This common microscopic origin constitutes
the correlation between the SMBH and Hubble-tension sectors in the
present framework; it does not imply a direct numerical relation
between the SMBH mass and the Hubble constant.
}
\textcolor{red}{
The same multi-axion framework can also accommodate a possible
connection to cosmic birefringence.  If the microscopic axions
couple to electromagnetism as
}
\be
\textcolor{red}{
 {\cal L}_{a\gamma}
 =
 -\frac{1}{4}
 \sum_i g_{i\gamma}a_i F_{\mu\nu}\widetilde F^{\mu\nu},
}
\label{agamma}
\ee
\textcolor{red}{
then a collective direction
$\phi_\alpha=\sum_i c_i^{(\alpha)}a_i$ inherits the effective coupling
}
\be
\textcolor{red}{
 g_{\alpha\gamma}
 =
 \sum_i c_i^{(\alpha)}g_{i\gamma}.
}
\label{geff}
\ee
\textcolor{red}{
The corresponding rotation of the polarization angle is schematically \cite{EskiltKomatsu2022,DiegoPalazuelos2022}
}
\be
\textcolor{red}{
 \beta
 =
 \frac{1}{2}
 \sum_\alpha g_{\alpha\gamma}
 \left[
 \phi_\alpha(t_0)-\phi_\alpha(t_{\rm LSS})
 \right],
}
\label{beta}
\ee
\textcolor{red}{
where $t_0$ denotes the present cosmic time and $t_{\rm LSS}$ the
cosmic time at the last-scattering surface.
Thus the collective direction $\phi_{28}$ responsible for the
pre-recombination modification of the expansion history could also
contribute to cosmic birefringence.  This provides an additional possible observational probe
of the same multi-axion field-space structure.  We do not, however,
regard cosmic birefringence as a prediction of the present framework,
since the coefficients $g_{i\gamma}$ depend on the microscopic
compactification and are not determined in the present analysis.
}
Whether a realistic heterotic compactification naturally realizes
the required alignment and harmonic relations is not assumed here
to be established.  Rather, Eq.~(\ref{EDEpotential}) represents the
effective condition that the underlying multi-axion sector must
approximately realize in order to produce a transient modification
of the pre-recombination expansion history.  This provides a
concrete connection between the string-axiverse structure and the
early-dark-energy mechanism relevant to the Hubble tension.

\section{Heterotic origin of the two cosmological
axion directions}
\label{heteroticconstraints}

The cosmological considerations in Sec.~3
lead to two qualitatively different
requirements on the axion potential.
The dark-matter direction requires
\be
V_{22}=\Lambda_{22}^4
\left(1-\cos\frac{\phi_{22}}{f_{22}}\right),
\ee
with
\be
m_{22}\sim 10^{-22}\ {\rm eV},
\qquad
f_{22}\sim 10^{16}\ {\rm GeV},
\ee
whereas the early-dark-energy direction
requires
\be
V_{28}=\Lambda_{28}^4
\left(1-\cos\frac{\phi_{28}}{f_{\rm eff}}
\right)^3,
\ee
with
\be
m_{28}^{\rm char}\sim 10^{-28}\ {\rm eV},
\qquad
f_{\rm eff}\sim 10^{18}\ {\rm GeV}.
\ee

The central question is whether these
two requirements can be satisfied within
a common heterotic compactification.

We address three aspects of this question:
the geometrical relation between instanton
actions and decay constants, the enhancement
of the effective decay constant through
alignment, and the simultaneous realization
of the two potential shapes and their
energy-scale hierarchy.
\subsection{K\"ahler moduli and axion kinetic terms}

Consider a heterotic compactification on a
Calabi--Yau threefold $Z$.
The K\"ahler form and the internal components of
the antisymmetric tensor field can be expanded
as
\be
 J=\sum_i t_i\omega_i,\qquad
 B_2=\frac{1}{2\pi}\sum_i b_i\omega_i,
\label{JB}
\ee
where $\omega_i$ form an integral basis of
$H^{1,1}(Z)$ and $b_i$ have period $2\pi$.
The normalization of $J$ is chosen such that
the internal volumes are measured in string units.

The real fields $t_i$ describe the K\"ahler
deformations of the internal manifold, whereas
$b_i$ correspond to the model-dependent axions
\cite{Svrcek,Leedom}.

In the four-dimensional ${\cal N}=1$ effective
theory, they combine into complexified K\"ahler
moduli,
\be
 T_i=t_i+i\frac{b_i}{2\pi}.
\label{complexmoduli}
\ee

The dimensionless internal volume is given by
\be
 {\cal V}
 =\frac{1}{6}\int_Z J\wedge J\wedge J
 =\frac{1}{6}\kappa_{ijk}t_it_jt_k,
\label{CYvolume}
\ee
where
\be
 \kappa_{ijk}=\int_Z
 \omega_i\wedge\omega_j\wedge\omega_k
\ee
are the triple intersection numbers.

At tree level and in the large-volume
approximation, the K\"ahler-moduli contribution
to the K\"ahler potential is
\be
 K=-\ln{\cal V}.
\label{Kpotential}
\ee
The other moduli are assumed to be fixed in
the following derivation.

The kinetic terms are
\be
 {\cal L}_{\rm kin}
 =-(M_{\rm Pl}^*)^2 K_{i\bar j}
 \partial_\mu T_i\partial^\mu\bar T_j,
\ee
where
\be
 K_{i\bar j}
 =\frac{\partial^2K}
 {\partial T_i\partial\bar T_j}.
\ee

The axion kinetic terms therefore take the form
\be
 {\cal L}_{\rm axion}
 =-\frac12 G_{ij}
 \partial_\mu b_i\partial^\mu b_j,
\ee
with
\be
 G_{ij}=\frac{(M_{\rm Pl}^*)^2}{2\pi^2}K_{i\bar j}.
\label{axionmetric}
\ee

Thus the normalization and mixing of the axions
are determined by the K\"ahler geometry.

\subsection{A geometrical constraint on instanton
actions and decay constants}

A worldsheet instanton wrapping a holomorphic
two-cycle $C_A$ is characterized by the integer
charges
\be
 Q_{Ai}=\int_{C_A}\omega_i.
\label{topologicalcharge}
\ee

Its classical action is
\be
 S_A=2\pi\int_{C_A}J
 =2\pi Q_{Ai}t_i.
\label{cycleaction}
\ee

The corresponding holomorphic exponential is
\be
 e^{-2\pi Q_{Ai}T_i}
 =e^{-S_A}e^{-iQ_{Ai}b_i}.
\label{instantonexp}
\ee

The same charge vector therefore determines
both the classical instanton action and the
axionic phase.

Introducing canonically normalized fields
\be
 {\boldsymbol\phi}=G^{1/2}{\boldsymbol b},
\ee
the axionic phase becomes
\be
 Q_A^T{\boldsymbol b}
 =Q_A^TG^{-1/2}{\boldsymbol\phi}.
\ee

We define the decay constant associated with
this individual instanton charge vector by
\be
 f_A^{-2}=Q_A^TG^{-1}Q_A.
\label{instantondecay}
\ee

Since ${\cal V}$ is a homogeneous function of
degree three, Euler's theorem gives
\be
 t_i\frac{\partial K}{\partial t_i}=-3.
\ee
Differentiating this relation with respect to
$t_j$, we obtain
\be
 t_it_j\frac{\partial^2K}
 {\partial t_i\partial t_j}=3.
\ee

Using
\be
 K_{i\bar j}
 =\frac14\frac{\partial^2K}
 {\partial t_i\partial t_j},
\ee
we find
\be
 t_iK_{i\bar j}t_j=\frac34,
\ee
and hence
\be
 t^TGt=\frac{3(M_{\rm Pl}^*)^2}{8\pi^2}.
\label{homogeneity}
\ee

The Cauchy--Schwarz inequality with respect
to the positive-definite metric $G$ implies
\be
 (Q_A^Tt)^2
 \leq
 (Q_A^TG^{-1}Q_A)(t^TGt).
\ee

Combining Eqs.~(\ref{cycleaction}),
(\ref{instantondecay}), and
(\ref{homogeneity}), we obtain \cite{Etheredge}
\be
 \boxed{
 S_A f_A\leq\sqrt{\frac32}M_{\rm Pl}^*.
 }
\label{actiondecaybound}
\ee

This inequality follows from the homogeneity
of the classical Calabi--Yau volume and does
not depend on the individual triple
intersection numbers.
Related constraints on axion decay constants
and instanton actions have been discussed
in the context of the axionic Weak Gravity
Conjecture \cite{Rudelius,Brown}.
Equality holds if and only if the charge
vector $Q_A$ is proportional to $Gt$.

Equation~(\ref{actiondecaybound}) establishes
a correlation between the instanton action
and the decay constant associated with an
individual worldsheet-instanton charge vector.
Its validity is restricted to the classical
K\"ahler geometry assumed above.

\subsection{Alignment and the cosmological hierarchy}

For $f_A=10^{16}$ GeV,
Eq.~(\ref{actiondecaybound}) allows instanton
actions as large as
\be
 S_A\lesssim294.
\ee
In contrast, for $f_A=10^{18}$ GeV, the same
inequality gives
\be
 S_A\lesssim2.94.
\ee

Consequently, a large decay constant and a
strong exponential suppression cannot be
simultaneously obtained from an individual
worldsheet-instanton charge vector within
the classical approximation.

This observation is particularly relevant
to the early-dark-energy direction considered
in this work, for which a large effective
decay constant and a small non-perturbative
scale are required.

The effective decay constant along a
collective direction must, however, be
distinguished from $f_A$ defined in
Eq.~(\ref{instantondecay}).

Let $c$ be a unit vector in the canonically
normalized axion field space.
Along the trajectory
\be
 {\boldsymbol\phi}=c\phi_{28},
 \qquad c^Tc=1,
\ee
the instanton phase is
\be
 Q_A^T{\boldsymbol b}
 =Q_A^TG^{-1/2}c\,\phi_{28}.
\ee

The projected decay constant is therefore
\be
 f_{A,c}^{-1}
 =\left|Q_A^TG^{-1/2}c\right|.
\label{projecteddecay}
\ee

In particular,
\be
 f_{A,c}\geq f_A.
\ee

Defining $\alpha_A$ as the angle between
$G^{-1/2}Q_A$ and $c$, we have
\be
 f_{A,c}=\frac{f_A}{|\cos\alpha_A|}.
\label{alignmentangle}
\ee

Equation~(\ref{actiondecaybound}) consequently
implies
\be
 S_A f_{A,c}|\cos\alpha_A|
 \leq\sqrt{\frac32}M_{\rm Pl}^*.
\label{alignmentbound}
\ee

Thus an enhanced effective decay constant
can coexist with a large instanton action
when the collective axion direction is
nearly orthogonal to the corresponding
canonically normalized charge vector.

For example, an instanton action of order
$100$ and a projected decay constant of
$10^{18}$ GeV require
\be
 |\cos\alpha_A|\lesssim 2.9\times10^{-2}.
\ee

The alignment mechanism therefore provides
a possible means of reconciling the two
requirements \cite{Ali}.
The existence of an appropriate collective
direction must be examined using the full
instanton charge matrix.

The characteristic potential scales required
by the two cosmological directions satisfy
\be
 \Lambda_{22}^4=m_{22}^2f_{22}^2,
 \qquad
 \Lambda_{28}^4=
 (m_{28}^{\rm char})^2f_{\rm eff}^2.
\ee

For the representative parameters considered
in this work, their ratio is
\be
 \frac{\Lambda_{22}^4}{\Lambda_{28}^4}
 \simeq10^8.
\label{cosmologicalhierarchy}
\ee

The realization of this hierarchy must be
consistent with both the instanton actions
and the kinetic geometry described above.

\subsection{Superpotential realization of the
sixth-order potential}

Worldsheet instantons can generate
non-perturbative superpotential contributions
of the form
\be
 W_{\rm np}
 =\sum_A{\cal A}_A
 e^{-2\pi Q_{Ai}T_i},
\label{heteroticW}
\ee
when the corresponding instanton zero-mode
conditions are satisfied
\cite{WittenWS,Curio,Beasley}.

The coefficients ${\cal A}_A$ depend on the
complex structure and bundle moduli.
The scalar potential must therefore be
derived from the full effective supergravity
rather than identified directly with the
individual instanton amplitudes.

In reduced Planck units, the $F$-term
potential is
\be
 V_F=e^K
 \left(
 K^{I\bar J}D_IW D_{\bar J}\overline W
 -3|W|^2
 \right).
\label{Fpotential}
\ee

An important observation concerning the
required early-dark-energy potential is
the identity
\be
 (1-\cos\theta)^3
 =\frac18\left|(1-e^{-i\theta})^3\right|^2.
\label{holomorphicidentity}
\ee

This identity suggests that the required
harmonic structure can be formulated in
terms of a third-order zero of a holomorphic
function.

To illustrate this possibility, consider
two effective complexified K\"ahler directions
\be
 T=t+i\frac{\theta_{22}}{2\pi},
 \qquad
 U=u+i\frac{\theta_{28}}{2\pi},
\ee
where
\be
 \theta_{22}=\frac{\phi_{22}}{f_{22}},
 \qquad
 \theta_{28}=\frac{\phi_{28}}{f_{\rm eff}}.
\ee

We introduce two stabilizer chiral multiplets
$X$ and $Y$ and consider the effective
superpotential
\be
 \begin{split}
 W={}&XA(e^{-2\pi T}-c_T)\\
 &+YB(e^{-2\pi U}-c_U)^3.
 \end{split}
\label{effectiveW}
\ee

Here $A,B,c_T,c_U$ are constants.
This superpotential is an effective-theory
ansatz, rather than a result derived from
a specified heterotic compactification.

We assume a K\"ahler potential of the form
\be
 K=K_0(T,\bar T,U,\bar U)
   +X\bar X+Y\bar Y+\cdots,
\ee
with canonical stabilizer metrics at
$X=Y=0$ and no kinetic mixing between
the stabilizers and the remaining moduli
at this point.

On the trajectory $X=Y=0$, and in the
absence of additional superpotential
contributions, we have
\be
 W=D_TW=D_UW=0,
\ee
whereas
\be
 D_XW=A(e^{-2\pi T}-c_T),
\ee
\be
 D_YW=B(e^{-2\pi U}-c_U)^3.
\ee

The resulting scalar potential is
\be
 \begin{split}
 V_F=e^{K_0}\big[
 &|A|^2|e^{-2\pi T}-c_T|^2\\
 +&|B|^2|e^{-2\pi U}-c_U|^6
 \big].
 \end{split}
\label{positivepotential}
\ee

Assuming that the real components are fixed
at $t=t_0$ and $u=u_0$, and choosing
\be
 c_T=e^{-2\pi t_0},
 \qquad
 c_U=e^{-2\pi u_0},
\ee
we obtain
\be
 \begin{split}
 V_F={}&\Lambda_{22}^4
 (1-\cos\theta_{22})\\
 &+\Lambda_{28}^4
 (1-\cos\theta_{28})^3,
 \end{split}
\label{combinedpotential}
\ee
where
\be
 \Lambda_{22}^4
 =2e^{K_0}|A|^2e^{-4\pi t_0},
\ee
and
\be
 \Lambda_{28}^4
 =8e^{K_0}|B|^2e^{-12\pi u_0}.
\ee

The two potential scales consequently obey
\be
 \boxed{
 \frac{\Lambda_{22}^4}{\Lambda_{28}^4}
 =
 \frac{|A|^2}{4|B|^2}
 e^{4\pi(3u_0-t_0)}.
 }
\label{commonhierarchy}
\ee

For equal coefficients $|A|=|B|$,
the cosmological hierarchy
(\ref{cosmologicalhierarchy}) requires
\be
 3u_0-t_0\simeq1.58.
\label{moduliconstraint}
\ee

The order of the holomorphic zero thus
enters both the local structure of the
potential and the hierarchy of its
characteristic scales.

However, the origin of the third-order
zero itself remains to be established.

Writing $z=e^{-2\pi U}$ and
\be
 F(z)=a_0+a_1z+a_2z^2+a_3z^3,
\ee
the conditions
\be
 F(z_0)=F'(z_0)=F''(z_0)=0,
 \qquad F'''(z_0)\ne0
\ee
imply
\be
 F(z)=a_3(z-z_0)^3.
\ee

Equivalently,
\be
 a_2=-3a_3z_0,\qquad
 a_1=3a_3z_0^2,\qquad
 a_0=-a_3z_0^3.
\label{tripleroot}
\ee

These relations translate the required
harmonic structure into constraints on
the holomorphic coefficients of the
effective superpotential.

\subsection{Conditions for a common heterotic
realization}

The preceding construction establishes
the compatibility of the two required
potential shapes at the level of an
effective supergravity ansatz.

A microscopic realization requires
additional conditions.

First, the effective directions $T$ and
$U$ must arise from a common K\"ahler metric
and an admissible instanton charge matrix.
In particular, the enhanced effective
decay constant must be consistent with
Eq.~(\ref{alignmentbound}).

Second, the relations
(\ref{tripleroot}) must follow from the
non-perturbative superpotential, rather
than being imposed independently on its
coefficients.

Third, the remaining moduli must be
stabilized without generating quadratic
and quartic terms along the early-dark-energy
direction.

The last condition is particularly important.
Writing
\be
 U=u_0+\delta u+i\frac{\theta_{28}}{2\pi},
\ee
the potential near the minimum behaves as
\be
 V_{28}\simeq
 e^{K_0}|B|^2e^{-12\pi u_0}
 \left[(2\pi\delta u)^2+\theta_{28}^2\right]^3.
\label{saxionpotential}
\ee

Thus the third-order holomorphic zero
eliminates the quadratic curvature not
only in the axion direction but also
in its saxionic partner.

An independent stabilization mechanism
is therefore required, and its backreaction
on the axion potential must be controlled.

The geometrical inequality
(\ref{actiondecaybound}), its extension
(\ref{alignmentbound}) to collective
directions, and the holomorphic conditions
(\ref{tripleroot}) provide complementary
constraints on a common heterotic origin
of the two cosmological axion directions.

\section{Discussion and conclusions}
\label{sec:conclusion}

In this work, we have investigated a unified
axion cosmology connecting grand unification,
dark matter, the formation of supermassive
black holes, and the Hubble tension.
Our purpose is not to introduce independent
axion-like particles for different cosmological
phenomena, but to investigate whether these
phenomena can be understood within a common
high-energy framework.

The QCD axion is associated with the minimal
supersymmetric SO(10) grand unified theory.
Its decay constant is related to the intermediate
symmetry-breaking scale, providing a connection
between the solution of the strong CP problem
and the structure of grand unification.
The two ultralight cosmological directions,
on the other hand, are considered within
the multi-axion sector of heterotic string
compactification.

The cosmological analysis in Sec.~3 leads
to two qualitatively different requirements
on this multi-axion sector.
The dark-matter direction $\phi_{22}$ requires
an ordinary axion potential,
\be
V_{22}=\Lambda_{22}^4
\left(1-\cos\frac{\phi_{22}}{f_{22}}\right),
\ee
with
\be
m_{22}\sim 10^{-22}\ {\rm eV},
\qquad
f_{22}\sim 10^{16}\ {\rm GeV}.
\ee
Its nonvanishing quadratic curvature allows
the field to behave as dark matter, while
its attractive self-interaction can induce
the instability of the Bose--Einstein
condensate and subsequent gravitational
collapse.
This mechanism provides a possible origin
of massive seeds for high-redshift
supermassive black holes.

In contrast, the early-dark-energy direction
$\phi_{28}$ requires a potential of the form
\be
V_{28}=\Lambda_{28}^4
\left(1-\cos\frac{\phi_{28}}{f_{\rm eff}}
\right)^3,
\ee
with the characteristic scales
\be
m_{28}^{\rm char}\sim 10^{-28}\ {\rm eV},
\qquad
f_{\rm eff}\sim 10^{18}\ {\rm GeV}.
\ee
The sixth-order behavior near the minimum
leads to an averaged equation of state
$\langle w\rangle=1/2$ in the regime of
rapid coherent oscillations.
The corresponding energy density decreases
as $a^{-9/2}$, faster than radiation.
This property allows the field to provide
a transient contribution to the expansion
rate near the matter--radiation transition.

The central question addressed in Sec.~4
is whether these two cosmological
requirements can be accommodated within
a common heterotic compactification.

In heterotic string theory, the axion
kinetic matrix and the worldsheet-instanton
actions are determined by the geometry
of the same internal manifold.
The geometrical relation
\be
S_A f_A\leq\sqrt{\frac{3}{2}}M_{\rm Pl}^*
\ee
expresses this correlation in the classical
large-volume approximation.
It shows that a large instanton action
and a large decay constant cannot be
independently assigned to an individual
instanton charge vector.

This constraint is particularly relevant
to the early-dark-energy direction,
which requires both a strongly suppressed
non-perturbative scale and a large
effective decay constant.
For a collective direction, however,
the corresponding relation becomes
\be
S_A f_{A,c}|\cos\alpha_A|
\leq\sqrt{\frac{3}{2}}M_{\rm Pl}^*.
\ee
Thus an appropriate alignment in the
multi-axion field space provides a
possible means of reconciling these
two requirements.

The remaining question concerns the
different functional forms of the two
cosmological potentials.
We have shown that these forms can be
obtained within a common effective
supergravity construction by introducing
the superpotential
\be
W=XA(e^{-2\pi T}-c_T)
+YB(e^{-2\pi U}-c_U)^3.
\ee
The first-order and third-order holomorphic
zeros generate quadratic and sixth-order
scalar potentials, respectively.

An important feature of this construction
is that the order of the holomorphic zero
enters not only the local structure of
the scalar potential but also its
characteristic energy scale.
The two potential scales satisfy
\be
\frac{\Lambda_{22}^4}{\Lambda_{28}^4}
=
\frac{|A|^2}{4|B|^2}
e^{4\pi(3u_0-t_0)}.
\ee
The cosmological requirement
\be
\frac{\Lambda_{22}^4}{\Lambda_{28}^4}
\sim 10^8
\ee
therefore imposes a quantitative condition
on the parameters of the effective
superpotential and the vacuum expectation
values of the K\"ahler moduli.
For equal coefficients $|A|=|B|$,
this condition reduces to
\be
3u_0-t_0\simeq 1.58.
\ee

The required sixth-order potential is
thereby reformulated as a condition on
the holomorphic structure of the
non-perturbative superpotential.
In particular, writing
$z=e^{-2\pi U}$, the relevant holomorphic
function must satisfy
\be
F(z_0)=F'(z_0)=F''(z_0)=0,
\qquad F'''(z_0)\neq 0.
\ee
The origin of this third-order zero
constitutes a concrete microscopic
problem for heterotic compactification.

A complete realization of the present
framework requires the simultaneous
satisfaction of three conditions.
First, the two cosmological directions
must arise from a common K\"ahler metric
and an admissible instanton charge matrix.
Second, the third-order holomorphic zero
must follow from the microscopic
non-perturbative dynamics rather than
from independently adjusted coefficients.
Third, the saxionic partners and the
remaining moduli must be stabilized
without generating significant quadratic
or quartic terms along the
early-dark-energy direction.

These conditions have not yet been
derived from a specific heterotic
compactification.
The superpotential considered here
should therefore be understood as an
effective construction that identifies
the microscopic requirements for a
common origin of the two cosmological
directions.

The same multi-axion structure may also
provide a connection to cosmic
birefringence through the electromagnetic
couplings of the collective axion
directions.
Such a connection remains dependent on
the microscopic anomaly coefficients
and is not a quantitative prediction
of the present analysis.

The main result of this work is thus
to translate two apparently distinct
cosmological requirements into a set
of geometrical and holomorphic
constraints on a common axion sector.
Rather than treating the axion decay
constants, instanton actions, and
potential shapes as independent
phenomenological inputs, we have
identified the relations that must
be satisfied in a heterotic
realization.

From this perspective, dark matter,
the formation of supermassive black
holes, and the Hubble tension provide
complementary probes of the geometry
and non-perturbative dynamics of the
underlying high-energy theory.
Cosmology can therefore serve as a
selection principle for string
compactification.

{\bf Acknowledgments}  

This work is partly supported by JSPS KAKENHI Grant Number 25H00653. 

\end{document}